\documentclass[twocolumn,aps,prl,floats,showpacs]{revtex4}
\newcommand{\beq}{\begin{eqnarray}}
\newcommand{\eeq}{\end{eqnarray}}

\usepackage{epsfig}
\usepackage{dcolumn}
\usepackage{bm}
\begin{document}

\title
{Landau theory of the Fermi-liquid to electron glass transition}
\author{Denis Dalidovich and V. Dobrosavljevi\'{c}}

\address{National High Field Magnetic Laboratory,\\
Florida State University, Tallahassee, Florida, 32310}

%
\begin{abstract}
A lattice model of spinless interacting electrons is used to
formulate the Landau theory of the Fermi liquid to electron glass
quantum phase transition. We demonstrate that the presence of
additional random site energies does not affect the character of
the transition, once the replica symmetry breaking is considered
self-consistently at the mean-field level. Inside the glass phase,
the low temperature conductivity assumes a non-Fermi liquid
$\delta\sigma\sim T^{3/2}$ form, in agreement with recent
experiments.
\end{abstract}
 \maketitle

Numerous recent experiments have demonstrated that the phase
diagram of the low carrier density systems proves remarkably rich
even in the absence of magnetic fields \cite{abrahams}. The
fascinating strong correlation physics has been observed whenever
the energy scale of the Coulomb repulsive interactions is
comparable to that of the Fermi energy. Once disorder is weak
enough, the interactions alone can lead to insulating behavior,
characterized by hard gaps in the charge sector.  But if the
magnitude of random potential is comparable to the Fermi energy,
electrons approach the strongly localized regime. In this case,
the interplay between localization and Coulomb interactions is
generally expected to lead to (gapless) glassy ordering of
electrons \cite{efros,dobr}.

Very interesting new information on this problem comes from recent
experiments by Bogdanovich and ${\rm Popovi\acute{c}}$
\cite{bogdan}. In this work, strong long-time fluctuations of
conductivity were observed as the electron density is reduced
below some critical value, indicating a dramatic slowing down of
electron dynamics. This behavior was attributed to the presence of
the glassy freezing that appears to come from the charge degrees
of freedom. Interestingly, the glassy behavior seems to emerge
appreciably before the metal-insulator transition, thus
identifying an intermediate metallic glass phase. Inside the glass
phase, unusual temperature dependence of the conductivity was
observed, which was fitted to the form $\delta\sigma\sim T^{3/2}$.

The main goal of the current Communication is to theoretically
examine the nature of the quantum phase transition from a Fermi
liquid to such a metallic glass phase, and the account for the
resulting modifications of charge transport. Previous works have
extensively examined the quantum paramagnet to spin-glass
transitions, regarding thus the spin degrees of freedom as the
only relevant ones at
criticality\cite{sach1,sach2,sengupta,rozenberg}. Here, we use a
similar approach to study the onset of glassiness in the charge
sector. Using a recently developed dynamical mean-field
formulation \cite{dobr}, we construct a Landau theory which
provides a complete description of the electron dynamics near the
relevant quantum critical point.

We consider a lattice model of spinless interacting electrons at
half-filling in the presence of on-site randomness, as given by
the Hamiltonian
\beq H=\sum_{ij} (-t_{ij}+\varepsilon_i
\delta_{ij})c_{i}^{\dagger}c_{j} +\sum_{ij} V_{ij}
c_{i}^{\dagger}c_{i}c_{j}^{\dagger}c_{j}.
\eeq
Here $t_{ij}$
denotes the corresponding hopping elements, while $V_{ij}$
describes the inter-site Coulomb interactions. The distribution of
the random site energies $\varepsilon_i$ is assumed to be
Gaussian, with variance $W$. As the simplest model \cite{dobr} for
the glassy freezing of electrons, we choose the inter-site
interactions $V_{ij}$ also to be Gaussian distributed random
\cite{dobrpast} variables, with variance $V$ . For this model, it
is straightforward \cite{dobr} to employ the methods of dynamical
mean-field theory \cite{kotl} which is formally exact in the limit
of large coordination. The averages over randomness are carried
out with using standard replica methods \cite{fishertz,dobr},
which are also used to identify the emergence of the glassy phase.
This procedure leads \cite{dobr} to the following single site
effective action ($\omega_m=2\pi (m+1)T$)
\beq\label{act1} &&
S_{{\rm eff}} (i) = \sum_{\omega_m} \sum_a c^{\dagger
a}_{i}(\omega_m)\left[ i\omega_m + t^2 G (\omega_m) \right]
c_{i}^a (\omega_m)\nonumber\\
&& -\int_0^{\beta} \int_0^{\beta} d\tau_1 d\tau_2
\left\{ \frac{V^2}{2}\sum_{ab} \delta n_i^a (\tau_1)
Q^{ab}(\tau_1 -\tau_2) \delta n_i^b (\tau_2)\right.\nonumber\\
&& + \left. \frac{W^2}{2} \sum_{ab}\delta n_i^a (\tau_1) \delta
n_i^b (\tau_2) \right\} \eeq In the formula above, the functional
integration is performed over the fermionic fields $c_i^a (\tau)$
($a=1,...,n$ are the standard replica indices). $\delta n_i^a
(\tau)=c_i^{\dagger a} (\tau)c_i^a (\tau)-\frac{1}{2}$ denotes the
deviation of the density from half-filling. We took into account
that the ``Weiss'' (cavity) field has the form $W_s
(\tau_1-\tau_2)=t^2 G(\tau_1-\tau_2)$ for electrons on a Bethe
lattice. The Green function $G(\tau_1-\tau_2)$ along with the
order parameter $Q^{ab}(\tau_1-\tau_2)$ must be determined
self-consistently using the effective action given by Eq.
(\ref{act1})\cite{dobr}. \beq\label{selfcon}
&G(\tau_1-\tau_2)=<c_i^{\dagger a} (\tau_1)c_i^a
(\tau_2)>_{\rm eff},&\\
\label{selfcon1} &Q^{ab}(\tau_1-\tau_2)=<\delta n_i^a (\tau_1)
\delta n_i^b (\tau_2)>_{\rm eff}.& \eeq The replica diagonal
components of $Q^{aa}(\tau)$ represent the averaged dynamic
compressibility, while the parameters $Q^{ab}$ ($a\ne b$) are time
independent and related to the familiar Edwards-Anderson (EA)
order parameter\cite{sach1}. Simple analysis of Eqs.
(\ref{act1})-(\ref{selfcon1}) shows that $q_{ab}$ is non-zero
everywhere, once $W\ne 0$. This is the consequence of the
non-uniform density due to the on-site random potential. As a
result, the glass transition for $W\ne 0$ assumes the character of
a De Almeida-Thouless line \cite{fishertz,dobr}, where special
care is needed in formulating a Landau theory. As it will be clear
from below, the relevant Landau theory can be rigorously
formulated, only if $W$ is small, but our conclusions remain
qualitatively correct for arbitrarily large $W$. If $V$ is much
smaller than $t$, action Eq. (\ref{act1}) describes the usual
disordered Fermi liquid\cite{dobrkot}, while in the opposite limit
the glassy ordering persists down to $T=0$\cite{dobr}. The quantum
phase transition to the glassy phase occurs in this model at some
critical value $(t/V)_{{\rm cr}}$, which has weak dependence on
$W$ \cite{dobr}.

To obtain the Landau functional we must perform a cumulant expansion
in Eq. (\ref{act1}) treating the term with
$Q^{ab}(\tau_1-\tau_2)$ as perturbation.
Before doing this, it is necessary to shift the $Q$-matrices,
eliminating thus the non-critical regular part in diagonal
elements ($\beta=1/T$):
\beq\label{shift}
Q^{ab}(\omega_n)\rightarrow Q^{ab}(\omega_n) -{\cal K}\delta^{ab}
-\beta \frac{W^2}{V^2}\delta_{\omega_n, 0}.
\eeq
Constant ${\cal K}$ must be formally determined from the condition of
absence of the term $\sum_{\omega_n}\sum_{ab}[Q^{ab}(\omega_n)]^{2}$
in the underlying Ginzburg-Landau action\cite{sach1}, that reads:
\beq\label{action}
&&\beta {\cal F}=\sum_{a,\omega_n} \left(
\frac{r+|\omega_n|}{V^2} \right)
Q^{aa}(\omega_n)+\frac{u}{2\beta}\sum_a \left[ \sum_{\omega_n}
Q^{aa}(\omega_n) \right]^2 \nonumber\\
&& -\frac{V^3}{3}\sum_{abc}\sum_{\omega_n} Q^{ab}(\omega_n)
Q^{bc}(\omega_n) Q^{ca}(\omega_n)
-\frac{\beta W^2}{2} Q^{ab}(\omega_n =0) \nonumber\\
&& -\frac{\beta y}{6} \int \int d\tau_1 d\tau_2 \sum_{ab}\left[
Q^{ab}(\tau_1-\tau_2) \right]^4.
\eeq
Here $r$, being some function of $t/V$, is the parameter that governs
the transition, while $u$ and $y$ are taken at $(t/V)_{{\rm cr}}$.
The presence of the last term, responsible for the RSB instability, is
crucial to further analysis. Accordingly, we employ the following
mean-field ansatz for the $Q$-matrices:
\beq\label{ansatz}
V^2 Q^{ab}(\omega_n)=
\left\{ \begin{array}{ll}
       \displaystyle D(\omega_n)+\beta q_{{\rm EA}}\delta_{\omega_n,0},
        & \quad a=b, \\
       \displaystyle \beta q_{ab}\delta_{\omega_n,0}, & \quad a\ne b.
       \end{array}\right.
\eeq In Eq. (\ref{ansatz}) $q_{{\rm EA}}$ is the EA order
parameter, and it is assumed that $q_{aa}=0$. The $\beta$-
prefactors are chosen to ensure the finite limit of the free
energy density as $T\rightarrow 0$. We must insert Eq.
(\ref{ansatz}) into the action Eq. (\ref{action}) and find the
saddle point solution with respect to the variations of $q_{ab}$,
$q_{{\rm EA}}$ and $D(\omega_n)$. $q_{{\rm EA}}$ and $q_{ab}$
should not, however, be varied independently \cite{georges}. They
must obey an additional relation between them that depends on the
presence of the glassy ordering and, hence, the replica symmetry
breaking. To make this point clear we first identify the part of
the action that contains only $q_{ab}$: \beq\label{partact} {\cal
F}_1=-R_1 {\rm Tr} q^2 -\frac{R_2}{3}{\rm Tr} q^3 -
\frac{R_3}{6}\sum_{a\ne b} (q_{ab})^4 - R_4 \sum_{a\ne b} q_{ab},
\eeq where, \beq\label{Rs}
\begin{array}{cc}
R_1=\beta (D(0)+\beta q_{{\rm EA}}), & R_2=\beta^2 \\ [2mm]
\displaystyle R_3=\frac{\beta y}{V^4}, & \displaystyle
R_4=\frac{\beta W^2 V^2}{2} .
\end {array}
\eeq

{\it Fermi liquid phase.} As we emphasized previously,
there is no replica symmetry breaking in the Fermi liquid phase.
Therefore, it is natural to choose the parametrization
$q_{ab}=q_{{\rm EA}}$ in this phase and take the variational
derivative of Eq. (\ref{partact}) with respect to $q_{{\rm EA}}$,
obtaining thus:
\beq\label{edwan}
2D(0)q_{{\rm EA}}+\frac{2y}{3V^4}q_{{\rm EA}}^3 +\frac{W^2 V^2}{2}=0
\eeq
Varying subsequently Eq. (\ref{action}) with respect
to $D(\omega_n)$ we arrive at the equation:
\beq\label{gap}
&& r+|\omega_n|+u \left[ \frac{1}{\beta} \sum_{\omega_n}
D(\omega_n) +q_{{\rm EA}}
\right] -D^2(\omega_n) \nonumber\\
&& -\frac{2y}{V^4} q_{{\rm EA}}^2 D(-\omega_n)
-\frac{2y}{V^4} \frac{q_{{\rm EA}}^2}{\beta} \sum_{\omega_1}
D(\omega_1)D(-\omega_1 -\omega_n) \nonumber\\
&& -\frac{2y}{3V^4} \frac{1}{\beta^2} \sum_{\omega_1, \omega_2}
D(\omega_1)D(\omega_2)D(-\omega_1 -\omega_2 -\omega_n)=0, \eeq
that closes the system of equations determining $q_{{\rm EA}}$ and
$D(\omega_n)$. All dangerous terms proportional to $\beta$ vanish
because we have judiciously chosen $q_{ab}$ equal to $q_{{\rm
EA}}$ from the very beginning. Note that Eq. (\ref{gap}) does not
contain $W$ at all, while $q_{{\rm EA}}=0$ only when $W=0$, as can
be seen from Eq. (\ref{edwan})

{\it Electron glass phase.} The saddle point solution of Eq.
(\ref{partact}) in this phase must be characterized by the Parisi
function $q(s)$ with $0\le s\le 1$. Considerations, that are
completely analogous to the classical
case\cite{sach1,fishertz,denphil}, lead to the functional form
that has two plateaus: \beq\label{parsol} q(s)= \left\{
\begin{array}{ll}
       \displaystyle q(0)=\left( \frac{3R_4}{4R_3} \right)^{\frac{1}{3}},
       \displaystyle 0<s<s_0=\left( \frac{6R_4 R_3^2}{R_2^3}
         \right)^{\frac{1}{3}} \\
       \displaystyle \frac{R_2 s}{2R_3},
       \displaystyle s_0 <s <s_1=1- \sqrt{1-\frac{4R_3R_1}{R_2^2}}\\
       \displaystyle q_{{\rm EA}}=
       \frac{R_2-(R_2^2-4R_1 R_3)^{1/2}}{2R_3}, s_1<s<1.
       \end{array}\right.
\eeq
The function $q(s)$ in the above equation saturates at the
value $q(1)=q_{{\rm EA}}$. This parametrization is in agreement
with the definition $q_{{\rm EA}}= {\rm max}_{a\ne b} q_{ab}$.
Combining Eqs. (\ref{Rs}) and (\ref{parsol}) we obtain that in the
glassy phase \beq\label{qglass} q_{{\rm EA}}^2=-[D(0)V^4]/y, \eeq
This equation connecting $D(0)$ and $q_{{\rm EA}}$ plays the same
role as Eq. (\ref{edwan}) in the Fermi liquid phase. We must
substitute then the solution Eq. (\ref{parsol}) into Eq.
(\ref{partact}) and add subsequently the result to the remaining
part of the action that contains $D(\omega_n)$. We will not write
down the expression for the free energy density obtained after
lengthy, but straightforward calculations, that are largely
identical to those done in the Appendix C of Ref. \cite{georges}.
We will state only that as a result of condition Eq.
(\ref{qglass}), there will be no terms in the free energy in which
$D(0)$ and $q_{{\rm EA}}$ are coupled explicitly. Taking the
variational derivative with respect to $D(\omega_n)$ results in
the same Eq. (\ref{gap}), albeit in the glassy phase $q_{{\rm
EA}}$ is connected with $D(0)$ by means of Eq. (\ref{qglass})
rather than Eq. (\ref{edwan}).

{\em Approaching the glass transition.} The necessity to obtain
$D(0)$ from Eq. (\ref{gap}) leaves us with the task of its
self-consistent solution. The exact analytical solution of this
non-linear integral equation is clearly out of question. However,
close to the $T=0$ transition point the {\it leading} order of the
correct solution is possible to obtain. The approximations we use
hinge also upon the smallness of $W$ and, consequently, $q_{{\rm
EA}}$.

We notice first that, if $y=0$, the complete solution is well-known
to be\cite{sach1,sach2,sengupta}
$D(\omega_n)=-\sqrt{|\omega_n|+\Delta}$, with $\Delta$ turning to zero
right at the critical point. Let's assume that for $y\ne 0$ the leading
approximation of $D(\omega_n)$ contains the same square root singularity
as for $y=0$, and analyze the role of the last two terms in
Eq. (\ref{gap}). The key point in this analysis is the value of the integral
\beq\label{integ}
J(\Omega_m)=T\sum_{\omega_n}\sqrt{|\omega_n|+\Delta}
\sqrt{|\omega_n +\Omega_m|+\Delta},
\eeq
that at $T=0$ is simply calculated to be:
\beq\label{J}
&&J(\Omega)=\frac{1}{2\pi} \left\{ \Lambda_{\omega}^{3/2}
-\frac{\Omega^2}{4}\ln \frac{2\Lambda_{\omega}}{|\Omega|}+\frac{\Omega^2}{2}
{\rm Arch} \frac{|\Omega|+2\Delta}{|\Omega|}\right. +\nonumber\\
&& \left. \left( \frac{|\Omega|}{2}+\Delta \right)^2 \arcsin
\frac{|\Omega|}{|\Omega|+2\Delta} -
(\Delta)^{3/2}\sqrt{|\Omega|+\Delta} \right\}. \eeq In Eq.
(\ref{J}) $\Lambda_{\omega}$ denotes the upper critical cutoff of
the order of unity. We see that all the terms in $J(\Omega)$,
except the first one proportional to $\Lambda_{\omega}^{3/2}$, are
of the order of $O(\Delta^2,\Omega^2)$. This means that the
prelast term in Eq. (\ref{gap}) gives contributions that depend
quadratically on small parameters $\Delta$ and $\omega_n$, and,
thus, subdominant to the leading terms that scale linearly with
them. The cutoff-dependent part in its turn leads to a mere
renormalization of the coefficient $u$ in $u q_{{\rm EA}}$. We
denote this renormalized term as ${\tilde u} q_{{\rm EA}}$.
Inserting then $J(\omega_1+\omega_n)$ to the last term of Eq.
(\ref{gap}) and integrating over $\omega_2$, we conclude, that the
only effect this term produces is to renormalize the critical
value $r_c$. This allows us to omit formally the last two
integrals in Eq. (\ref{gap}) and resolve the ensuing quadratic
equation for $D(\omega_n)$, obtaining that in the Fermi liquid
phase: \beq\label{D} D(\omega_n)= -\frac{y q_{{\rm
EA}}^2}{V^4}-\sqrt{|\omega_n|+\Delta}, \eeq \beq \Delta=r-r_c
+{\tilde u}q_{{\rm EA}} \eeq in the leading approximation. It is
easily verifiable $\acute{a}$ {\it posteriori} that the first term
in Eq. (\ref{D}), being inserted into the last two integrals of
Eq. (\ref{gap}), renders the contributions of the higher order of
smallness compared to ${\tilde u}q_{{\rm EA}}$. Together Eqs.
(\ref{edwan}) and (\ref{D}) determine completely the $T=0$
behavior of the disordered phase near the quantum critical point.
As as a result of their solution, one can distinguish the
following regimes on a ($r-r_c$, $W$) plane, schematically
depicted on Fig. (1).

(I) In this regime, in which
$W \ll (r-r_c)^{3/4}$ and can be treated as a perturbation,
$q_{{\rm EA}}=(W^2 V^2)/4\sqrt{r-r_c}$,
$\Delta \approx r-r_c$.

(II) This region is characterized by $|r-r_c|^{3/4} \ll W$.
As a result, we have
$q_{{\rm EA}}\approx (W^2 V^2 /4\sqrt{{\tilde u}})^{2/3}$
and $\Delta \approx {\tilde u}q_{{\rm EA}}$ in this regime.

(III) This regime,
in which $(r_c-r)^{3/4} \gg W$, is the closest
to the $T=0$ critical boundary. EA order parameter,
that crosses over to its value in the glassy phase, is given by
$q_{{\rm EA}}=[(r_c -r)/{\tilde u}]+(\Delta /{\tilde u})$, with
\beq\label{delnear}
\Delta =\left( \frac{2y (r_c -r)^2}{3{\tilde u}^2 V^4}-
\frac{W^2 V^2 {\tilde u}}{4(r_c -r)} \right)^2.
\eeq
From Eq. (\ref{delnear}) it is easily seen that $\Delta$ vanishes at the
critical line given by
$W=(8y/3)\left[ (r_c -r)/{\tilde u}V^2 \right]^{3/2}$.
\begin{figure}
\begin{center}
\epsfig{file=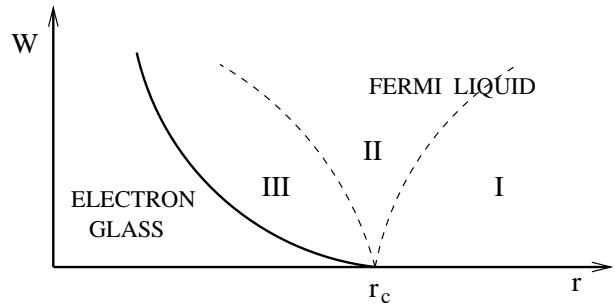, height=4cm}
\caption{Zero-temperature phase diagram demarcating the regions
of Fermi liquid and electron glass phases. For small $W$, $r-r_c$ the
curve separating these two phases scales roughly as $W\sim (r_c-r)^{3/2}$.
The limiting behavior of the EA order parameter
$q_{{\rm EA}}$ and gap $\Delta$ in different regimes is presented in the
text.}
\label{1}
\end{center}
\end{figure}
This is exactly the line that determines the transition to the
glassy phase obtained from the joint solution of Eqs.
(\ref{edwan}), (\ref{gap}) and (\ref{qglass}). Taking into account
Eq. (\ref{qglass}), we resolve similarly the equation for
$D(\omega_n)$ to get that below this line
\beq\label{dglass}
D(\omega_n)= -\frac{y q_{{\rm EA}}^2}{V^4} -\sqrt{|\omega_n|},\quad
q_{{\rm EA}}=\frac{r_c-r}{{\tilde u}}.
\eeq
Within the spin glass
phase, that obtains for all values of $W$, $\Delta$ is zero
everywhere. Therefore, the imaginary part of the local dynamic
susceptibility has the non-Fermi liquid singularity of the form
$\sim {\rm sgn}(\omega) \omega^{1/2}$. The transition to the
electron glass phase occurs as a second-order transition and is of
the same character for both zero and non-zero on-site randomness
$W$.

{\em Finite temperature behavior.} The evaluation of the
temperature-dependent correction to the integral $J(\Omega_m)$ in
Eq. (\ref{integ}) leads to the following results. If $\Delta \ll
T$, the correction scales $\propto T^2$ for $|\Omega_m| \sim T$
and $\propto T^{3/2}|\Omega_m|$ for $|\Omega_m|\gg T$. If $\Delta
\gg T$, it behaves $\propto T^2$ for $|\Omega_m| \sim T$ and
$\propto T^2|\Omega_m| /\sqrt{\Delta}$ for $|\Omega_m| \gg T$. As
a result, the equations governing the dependence of parameters
$q_{{\rm EA}}(T)$ and $\Delta (T)$ at finite temperature can be
obtained from those considered above by the formal substitution of
$r_c$ by $r_c(T)=r_c -c_1 T^2/\sqrt{\Delta}$ for $\Delta \gg T$,
and $r_c(T)=r_c -c_2 T^{3/2}$ in the opposite limit. $c_1$ and
$c_2$ are now, however, some complicated cutoff-dependent
coefficients of the order of unity. Those modifications lead to
the myriad of limiting cases close the quantum critical point that
will not be enumerated here. Instead, we mention that only if $T$
is small enough in the Fermi liquid phase, all temperature
corrections behave as $T^2$ with some large prefactors arising
because $W,|r-r_c| \ll 1$. On the contrary, for all temperatures
in the glassy phase $q_{{\rm EA}}(T)=[r_c -r -c_2 T^{3/2}]/{\tilde
u}$. The calculation of the electron self-energy\cite{sengupta} in
the lowest order using Eqs. (\ref{shift}), (\ref{ansatz}) and
(\ref{dglass}) suggests, that the leading temperature correction
to the elastic scattering rate is proportional to $T^{3/2}$ in the
glassy phase. This gives rise to the non-Fermi liquid temperature
dependence of the conductivity, $\delta \sigma \sim T^{3/2}$, in
qualitative agreement with the experimental observations
\cite{bogdan}.

We emphasize that our results are based on the non-perturbative
treatment of the RSB term in Eq. (\ref{action}) and the specific
conditions, connecting $q_{{\rm EA}}$ and $q_{ab}$ \cite{georges}.
An alternative saddle point solution of Eq. (\ref{action}), in
which $q_{{\rm EA}}$ and $q_{ab}$ are allowed to be varied
independently \cite{sach1}, leads the gapless glassy phase, only
if $W=0$, which we believe is incorrect. Our solution which has
zero gap for {\it all} finite $W$, seems to be the one in
agreement with the recent numerical simulations \cite{arrachea}.
Though our theory is formulated for small $W$,  the saddle-point
Eqs. (\ref{gap}) and (\ref{qglass}) are free from this parameter.
This strongly suggests that the similar set of equations, with
renormalized coefficients, should describe the glassy phase close
to the transition for all $W$, including the limit $W\rightarrow
\infty$ \cite{dobr}.

To conclude, we have presented a Landau theory description of the
disordered Fermi liquid to electron glass quantum critical
behavior. Our results represent an exact solution \cite{miller1}
of the model within a dynamical mean-field formulation, which is
formally exact in the limit of large coordination. It is important
to note that a glass transition having a character of a De
Almeida-Thouless line \cite{fishertz}, such as the one we
describe, generally emerges within mean-field models. An
alternative formulation \cite{droplet}, based on droplet
approaches predicts the absence of such transitions for models
with short-range interactions. In the case of an electron glass,
the existence of the long-range Coulomb interaction opens a
possibility that droplet approaches are not relevant, and that the
glassy behavior of electrons could be well described using
mean-field models. This possibility seems to find support in very
recent experiments \cite{jan}, which provide striking evidence of
scale-invariant dynamical correlations inside the glass phase,
consistent with the hierarchical picture of glassy dynamics as
emerging from mean-field models.

We thank L. Arrachea, J. Jaroszynski, D. Popovi\'c, M. Rozenberg,
and especially S. Sachdev  for useful discussions. This work was
supported by the National High Magnetic Field Laboratory (DD and
VD), and the NSF grant DMR-9974311 (VD).


\begin{thebibliography}{99}
\bibitem{abrahams} E. Abrahams, S. V. Kravchenko, M. P. Sarachik,
Rev. Mod. Phys. {\bf 73}, 251 (2001).
\bibitem{efros} A. L. Efros {\it et al.}, J. Phys. C {\bf 8} L49 (1975).
\bibitem{dobr} A. A. Pastor and V. ${\rm Dobrosavljevi\acute{c}}$,
Phys. Rev. Lett. {\bf 83}, 4642 (1999).
\bibitem{bogdan}S. Bogdanovich and D. ${\rm Popovi\acute{c}}$,
Phys. Rev. Lett. (2002, in press), cond-mat/0106545.
\bibitem{sach1} N. Read, S. Sachdev, J. Ye, Phys. Rev. B {\bf 52}, 384 (1995).
\bibitem{sach2} S. Sachdev, N. Read, R. Oppermann,
Phys. Rev. B {\bf 52}, 10286 (1995).
\bibitem{sengupta} A. Sengupta, A. Georges, Phys. Rev. B {\bf 52},
10295 (1995).
\bibitem {rozenberg}M. J. Rozenberg and D. R. Grempel Phys. Rev. Lett.
\textbf{81}, 2550 (1998); D. R. Grempel and M. J. Rozenberg,
Phys. Rev. Lett. \textbf{80}, 389 (1998); L. Arrachea and M. J.
Rozenberg, Phys. Rev. Lett. \textbf{86}, 5172 (2001).
\bibitem{dobrpast} Recent work \cite{dobr} has shown that
in presence of random site energies, such random interactions are
generated by renormalization even if one starts with purely
repulsive (e. g. Coulomb) interactions in the bare Hamiltonian.
Similar conclusions have been obtained numerically in: E. R.
Grannan and C. C. Yu, Phys. Rev. Lett. {\bf 71}, 3335 (1993).
\bibitem{kotl} A. Georges, G. Kotliar, W. Krauth and M. J. Rosenberg,
Rev. Mod. Phys. {\bf 68}, 13 (1996).
\bibitem{fishertz} K. H. Fischer and J. A. Hertz, {\it Spin Glasses}
(Cambridge University Press, Cambridge, England, 1993).
\bibitem{georges} A. Georges, O. Parcollet, S. Sachdev, Phys. Rev. B
{\bf 63}, 134406 (2001).
\bibitem{dobrkot} V. ${\rm Dobrosavljevi\acute{c}}$ and G. Kotliar,
Phys. Rev. B {\bf 50}, 1430 (1994).
\bibitem{sach3}S. Sachdev and N. Read, J. Phys. Cond. Matt. {\bf 8},
9723, (1996)
\bibitem{denphil} Denis Dalidovich and Philip Phillips, Phys. Rev. B
{\bf 59}, 11925 (1999).
\bibitem{arrachea} L. Arrachea, M. Rozenberg, private communication.
The simulations were performed for the random Ising model in both
transverse and longitudinal magnetic fields. The behavior near the
quantum critical point is governed by the same action Eq.
(\ref{action}), with $|\omega_n|$ substituted by $\omega_n^2$.
\bibitem{miller} J. Miller and  D. Huse, Phys. Rev. Lett.,
{\bf 70}, 3147, (1993).
\bibitem{miller1}  To independently verify the validity of our
Landau theory solution, we have also used an alternative approach,
following Ref. \cite{miller}. By explicit computation, we have
shown that the loop corrections to the irreducible polarizabilty
(see Ref. \cite{miller}), represent only sub-leading
contributions, so that the predictions of the Landau theory
provide an exact description of the quantum critical behavior
within our mean-field model.
\bibitem{droplet} D. S. Fisher and D. Huse, Phys. Rev. B
{\bf 38}, 373 nd 378 (1988).
\bibitem{jan} J. Jaroszynski, Dragana Popovi\'c, and T. M.
Klapwijk, cond-mat/0205226

\end{thebibliography}
\end{document}